\documentclass{aa}
\usepackage{graphicx}
\def\ls{\mathrel{\lower0.6ex\hbox{$\buildrel {\textstyle <}
 \over {\scriptstyle \sim}$}}}
\def\gs{\mathrel{\lower0.6ex\hbox{$\buildrel {\textstyle >}
 \over {\scriptstyle \sim}$}}}
\begin{document}
%
   \title{The extended counterpart of submm source Lockman850.1
   \thanks{Based on Observations carried out with the IRAM Plateau de
    Bure Interferometer. IRAM is supported by INSU/CNRS (France),
    MPG (Germany) and ING (Spain).}}


   \author{D.~Lutz
          \inst{1}
          \and
          J.S.~Dunlop
          \inst{2}
          \and
          O.~Almaini
          \inst{2}
          \and
          P.~Andreani
          \inst{6,1}
          \and
          A.~Blain
          \inst{8}
          \and
          A.~Efstathiou
          \inst{4}
          \and
          M.~Fox
          \inst{4}
          \and
          R.~Genzel
          \inst{1}
          \and
          G.~Hasinger
          \inst{9,1}
          \and
          D.~Hughes
          \inst{5}
          \and
          R.J.~Ivison
          \inst{3}
          \and
          A.~Lawrence
          \inst{2}
          \and
          R.G.~Mann
          \inst{2}
          \and
          S.~Oliver
          \inst{7}
          \and
          J.A.~Peacock
          \inst{2}
          \and
          D.~Rigopoulou
          \inst{1}
          \and
          M.~Rowan-Robinson
          \inst{4}
          \and
          S.~Scott
          \inst{2}
          \and
          S.~Serjeant
          \inst{10}
          \and
          L.~Tacconi
          \inst{1}
          }

   \offprints{D. Lutz}

   \institute{Max-Planck-Institut f\"ur extraterrestrische Physik,
              Postfach 1312, 85741 Garching, Germany\\
              email: lutz@mpe.mpg.de
         \and 
             Institute for Astronomy, Department of Physics \& Astronomy,
             University of Edinburgh, Blackford Hill, Edinburgh EH9 3HJ, UK
         \and
              UK ATC, Royal Observatory, Blackford Hill, Edinburgh EH9 3HJ, UK 
         \and
             Astrophysics Group, Imperial College London, Blackett Laboratory,
             Prince Consort Road, London SW7 2BZ, UK
         \and
             Instituto Nacional de Astrofisica, Optica y Electronica (INAOE),
             Apartado Postal 51 y 216, 72000 Puebla, Pue., Mexico
         \and
             Osservatorio Astronomico di Padova, vicolo 
             dell'Osservatorio 5, 35122 Padova, Italy
         \and 
             Astronomy Centre, University of Sussex, Falmer, Brighton BN1 9QJ, UK
         \and
             Institute of Astronomy, University of Cambridge, 
             Madingley Road, Cambridge CB3 0HA, UK
         \and
             AIP, An der Sternwarte 16, 14482 Potsdam, Germany 
         \and
             Unit for Space Sciences and Astrophysics, University of Kent,
             Canterbury, Kent CT2 7NR, UK
             }

   \date{Received; accepted}

   \abstract{The IRAM Plateau de Bure mm interferometer and 
deep K-band imaging have been used to identify
the brightest submm source detected in the Lockman field of the UK 8mJy SCUBA
survey. The near infrared counterpart is an extended (20--30\,kpc), clumpy, 
and extremely red object. The spectral energy distribution suggests it to be 
a dusty star forming object at a redshift of about 3 (2--4). Its star 
formation rate and near-infrared properties are 
consistent with Lockman850.1 being a massive elliptical in formation.
            }

\maketitle
%

\section{Introduction}

Over the past years, detections of (sub)mm sources in deep surveys have 
opened a new window on the high redshift universe (e.g. Smail, Ivison
\& Blain \cite{smail97}, Hughes et al. \cite{hughes98}, Barger et al.
\cite{barger98}, Eales et al. \cite{eales99}, Bertoldi et al. 
\cite{bertoldi00}).
The source population discovered by SCUBA contributes a large fraction 
of the cosmic submm background.
A significant part of star formation at redshifts 2 to 4 may be located 
in such dusty and extremely luminous galaxies. 
With inferred star formation rates of $\approx$1000\,M$_\odot$/yr, they may 
indicate the formation of present day massive spheroids. 
A direct link between star formation/spheroid growth on one side and 
AGN fuelling/black hole growth on the other is not only plausible through
gas fuelling triggered by interactions or merging, but
also suggested by the correlation between black hole and spheroid
mass (Magorrian et al. \cite{magorrian98}) and by global similarities in
the cosmic evolution of star formation and quasar activity. In turn,
this relates to the question of the starburst and AGN contribution to
the power output of SCUBA sources, and their role in the origin of the
cosmic X-ray and infrared backgrounds (e.g. Fabian et al. \cite{fabian00},
Hornschemeier et al. \cite{hornschemeier00})

\begin{figure*}
\center{\resizebox{13.cm}{!}{\rotatebox{-90}{\includegraphics[0,18][320,601]
{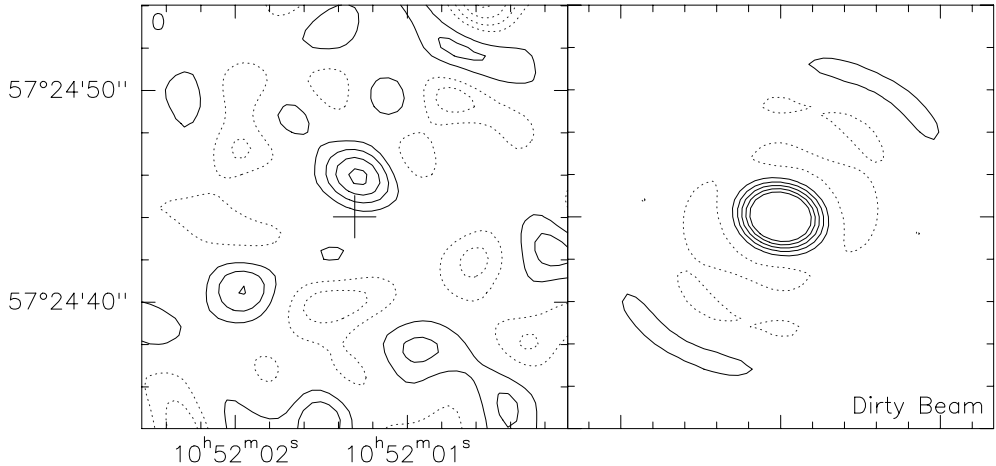}}}}
\caption{IRAM Plateau de Bure 1.26mm interferometric map of Lockman850.1 (left,
not CLEANed) along with the corresponding dirty beam including the
sidelobes (right). Linear contours
are spaced by 0.7mJy. The cross denotes the phase center of the observation.}
\label{fig:pdbmap}
\end{figure*}

Progress in all these issues relies on accurate identifications and
redshift determinations, which are extremely difficult because of
the large (15\arcsec\/) SCUBA beam and the
high density of very faint possible optical counterparts. 
A key step in this identification process is accurate astrometry. This 
can be obtained through mm interferometry, which directly locates the 
thermal dust emission (Downes et al. \cite{downes99}, Bertoldi et al.
\cite{bertoldi00}, Gear et al. \cite{gear00}, Frayer et al. \cite{frayer00}). 
Such astrometry is demanded by the variance in 
optical/near-infrared properties of proposed counterparts, which
makes color-based identification criteria indecisive. Colors range
from fairly blue (e.g. SMM02399, Ivison et al. \cite{ivison98}) to fainter 
very red objects (e.g. Smail et al. \cite{smail99}, 
Gear et al. \cite{gear00}, Bertoldi et al. \cite{bertoldi00}) 
and may include objects that are optically blank to the limit
of the deepest HST images (Smail et al. \cite{smail00},
Muxlow et al. \cite{muxlow01}). Accurate source positions and identifications also 
test hypotheses suggesting a local nature of part
of the submm sources (Lawrence \cite{lawrence01}).

The UK submm survey consortium is undertaking a wide field survey of
850$\mu$m sources with 3-sigma flux densities $>$8\,mJy. This survey is centred
on the Lockman Hole (East), and the ELAIS N2 field, both previously
surveyed with ISOPHOT at 175$\mu$m (Kawara et al. \cite{kawara00}, 
Lagache et al. \cite{lagache01}) and with the VLA (de Ruiter et al. 
\cite{ruiter97}, Ciliegi et al. \cite{ciliegi99}). Spectral energy 
distribution (SED) arguments 
suggest that SCUBA sources detected neither by ISOPHOT nor in the moderately
deep VLA data must lie at $z\gs 1.5$. Lockman850.1 is the brightest
SCUBA source detected in the Lockman Hole part of this survey (Scott et
al. \cite{scott01}), a region extensively observed in X-rays 
(Hasinger et al. \cite{hasinger98}, \cite{hasinger01}).

\section{Millimeter interferometry}

In order to accurately locate the (sub)mm emission of Lockman850.1, 
the source was observed with the IRAM Plateau de
Bure (PdB) interferometer (Guilloteau et al. \cite{guilloteau92}) for 
a total of 25.9 hours on 1999 June 22, August 30,
October 31, and November 5. Observations were done partly in the 5D 
and partly in the 4D configurations,
with phase center at the nominal position from the SCUBA map 
(Table~\ref{tab:positions}). Receivers were tuned to 238.46 GHz (1.26mm)
and 90.0 GHz (3.33mm).
The flux scale was calibrated using observations of 3C273 and 0923+392. 
Temporal 
variations of amplitude and phase were calibrated by frequent observations of
0917+624 and 1044+719. 
Data were reduced at IRAM Grenoble using GAG software. A positive signal is
seen in all individual tracks. The final map shown in Fig.~\ref{fig:pdbmap} 
was produced with natural weighting and has not been CLEANed.

We derived the total flux at 238 GHz of $3.03\pm 0.56$\,mJy, and the position 
listed in 
Table~\ref{tab:positions} from a point source fit to the visibilities. The 
positional uncertainty is obtained from adding in quadrature the statistical
error of the fit and an astrometric uncertainty of 0.2\arcsec\ (see Downes et 
al. \cite{downes99}). At 
the signal-to-noise of our data, no strong constraint can be put on the 
size of the mm source --
the 1.26\,mm map is consistent with the 
2.8\arcsec$\times$2.0\arcsec\ FHWM beam of our observations as well as with
a centrally peaked structure extended on a few arcsec scale.
No source was detected in the 3mm band (3$\sigma$ upper limit 0.6\,mJy at 
90\,GHz), in agreement with the emission being due to dust.

We have obtained additional photometry of Lockman850.1 on 2000 March 2 
and March 5 using the MAMBO bolometer array at the IRAM 30m telescope in 
on-off mode. The 11\arcsec\ beam was 
centered on the PdB interferometric position (Table~\ref{tab:positions}). 
Averaging all 45 minutes of data we obtain a 1.2\,mm flux of $3.8\pm 0.5$\,mJy.
In addition to the detection map, 850$\mu$m photometry was obtained with SCUBA
in January 2000 confirming the detection with a flux of 10.3$\pm$2.1\,mJy. 
A marginal detection is obtained from the SCUBA map at 450$\mu$m.
We adopt the mean of two independent reductions 
(Fox et al. \cite{fox01}, Scott et al. \cite{scott01}).
Photometry of Lockman850.1 is completed by a detection at 
1.4GHz (Ivison et al. \cite{ivison01}). The
position of the radio detection is in excellent agreement with the mm
detection (Table~\ref{tab:positions}). In the 5GHz data of Ciliegi et al.
(\cite{ciliegi01}) a $2.5\sigma$ maximum is seen, consistent with extrapolation
from the
1.4GHz point assuming nonthermal emission. We list a 5GHz upper limit
in Table~\ref{tab:photom}.

\section{Optical/near-infrared imaging}

\begin{figure}
\resizebox{8.8cm}{!}{\includegraphics{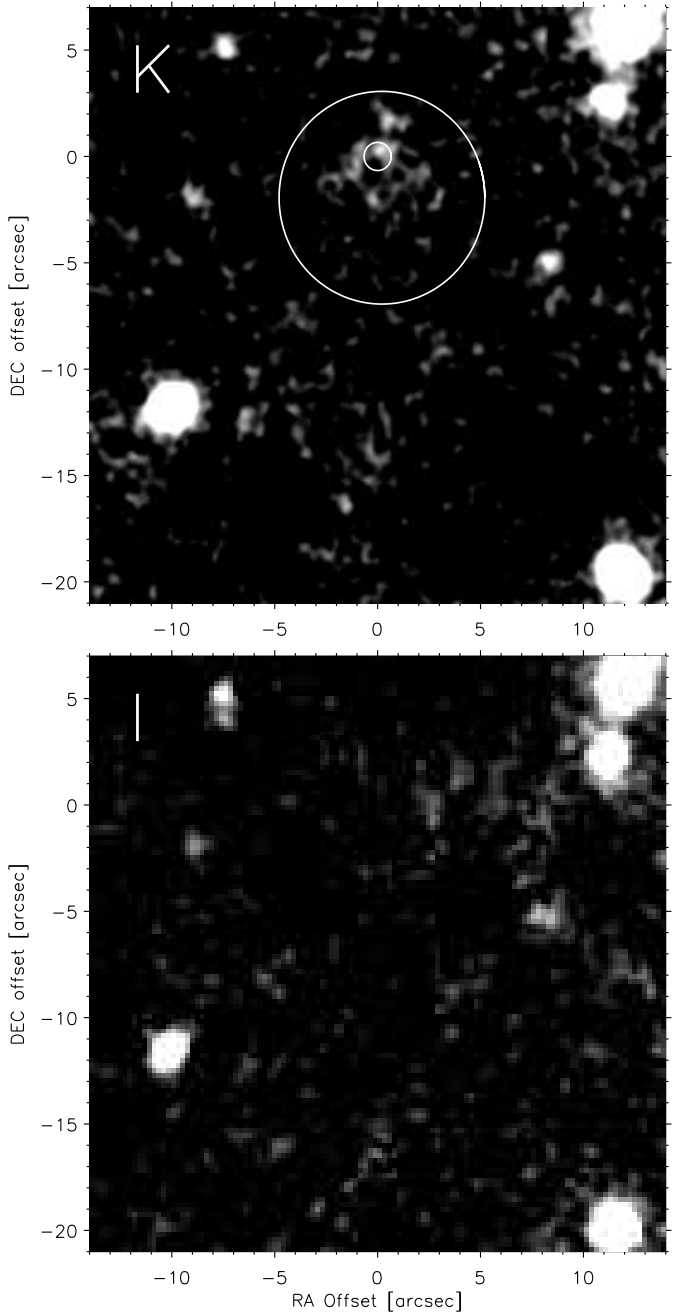}}
\caption{Top: K band image of the submm source Lockman850.1. The big circle 
denotes a 5\arcsec\ uncertainty radius around the nominal SCUBA 
position, the small circle a 0.66\arcsec\ (2$\sigma$) uncertainty radius 
around the IRAM interferometric position. Offsets in arcsec are given with
respect to the IRAM position. Bottom: I band image of the same
region, showing the nondetection of Lockman850.1 at that wavelength.
}
\label{fig:kiimages}
\end{figure}

A deep K band image of Lockman850.1 was obtained in 246 minutes total 
integration time during the nights of May 17-21, 2000, using the UKIRT 
Fast Track Imager. The K-band seeing in the coadded data is 0.9\arcsec.
A knotty extended object is clearly detected at the position of the
submm source (Fig.~\ref{fig:kiimages}, Table~\ref{tab:positions}), we 
list in Table~\ref{tab:photom} photometry for both the central knot and 
an aperture enclosing the extended K band emission. No optical counterpart
is detected in a deep I-band image taken at the WHT Prime Focus Imager
(3500sec, 0.75\arcsec\ seeing, see Fig.~\ref{fig:kiimages}) and in less deep 
R and V imagery. At I$-$K$>$6.2, Lockman850.1
is a bona fide extremely red object (ERO), and a Class I SCUBA source
in the terminology of Ivison et al. (\cite{ivison00a}).

\begin{table}
\begin{tabular}{lll}\hline
             &RA (J2000)        & DEC (J2000)   \\ \hline
SCUBA        &10 52 01.31$\pm$2.5\arcsec &+57 24 44.0$\pm$2.5\arcsec\\
IRAM PdB     &10 52 01.284$\pm$0.33\arcsec&+57 24 45.94$\pm$0.28\arcsec\\
K-band peak  &10 52 01.26$\pm$0.3\arcsec&+57 24 46.2$\pm$0.3\arcsec\\
VLA 1.4GHz   &10 52 01.252$\pm$0.05\arcsec&+57 24 45.86$\pm$0.05\arcsec\\ \hline
\end{tabular}
\caption{Position of Lockman850.1 (J2000).
The positional errors do not 
include systematic offsets between the radio and
optical (USNO A2) reference frame estimated at $\approx0.3$\arcsec.}
\label{tab:positions}
\end{table}

\begin{table}
\begin{tabular}{lrr}\hline
Band             &Flux/Magnitude       & Note \\ \hline
0.5-2keV         &$<3\,10^{-16}$erg\,s$^{-1}$\,cm$^{-2}\hspace{-5.5mm}$&
Hasinger et al. \cite{hasinger01} \\
m$_I$            &     $>$26.0mag&in 4\arcsec diameter\\
m$_I$            &     $>$27.5mag&in 1\arcsec diameter\\
m$_K$            &19.8$\pm$0.2mag&in 4\arcsec diameter\\
m$_K$            &21.4$\pm$0.05mag&in 1\arcsec diameter\\
S$_{7\mu m}$     &       $<$0.2mJy&Fadda et al. \cite{fadda01}\\
S$_{15\mu m}$    &       $<$0.2mJy&Fadda et al. \cite{fadda01}\\
S$_{175\mu m}$   &       $<$200mJy&Estim. from Kawara \cite{kawara00}\\
S$_{450\mu m}$  &      35.0$\pm$10mJy&SCUBA 8\arcsec\ beam\\
S$_{850\mu m}$  &      10.5$\pm$1.6mJy&SCUBA 15\arcsec\ beam\\
S$_{1.2mm}$      &3.8$\pm$0.5mJy&MAMBO 11\arcsec\ beam\\ 
S$_{1.26mm}$     &3.03$\pm$0.56mJy&PdB interferometer\\
S$_{3.33mm}$     &   $<$0.6mJy&PdB interferometer\\
S$_{5GHz}$       &   $<$0.07mJy&Ciliegi et al. \cite{ciliegi01}\\
S$_{1.4GHz}$     &0.074$\pm$0.013mJy&Ivison et al. \cite{ivison01}\\ \hline
\end{tabular}
\caption{Photometry of Lockman850.1}
\label{tab:photom}
\end{table}

\section{Discussion}

\begin{figure*}
\center{\resizebox{14.cm}{!}{\includegraphics{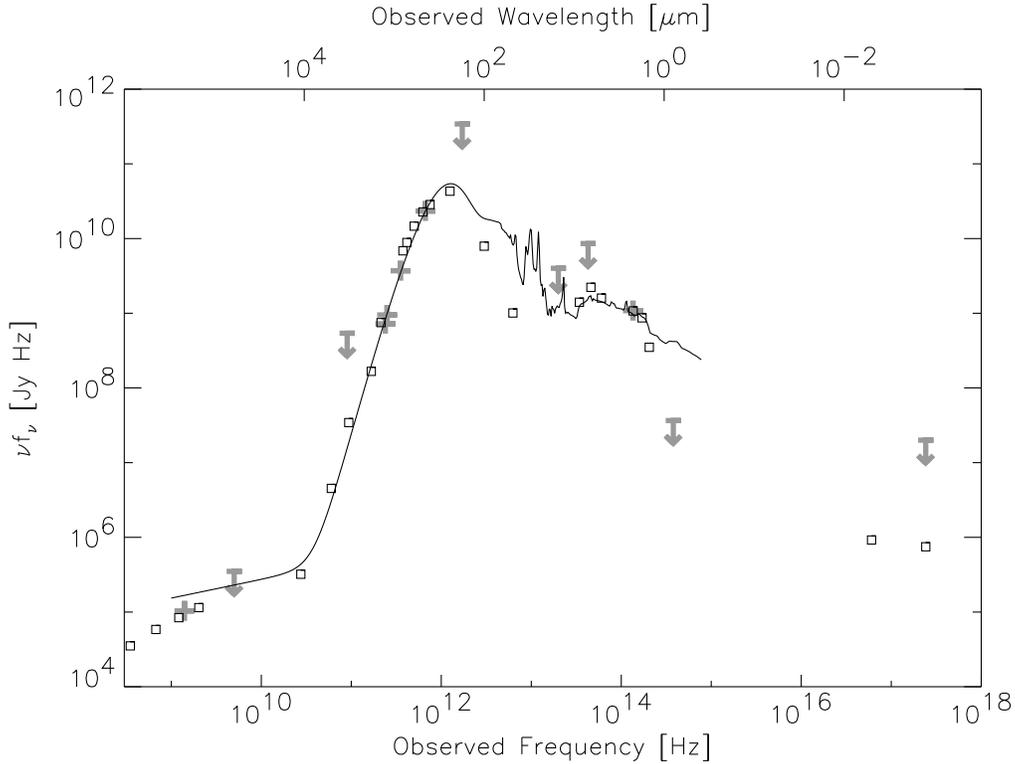}}}
\caption{Spectral energy distribution of Lockman850.1 (thick grey crosses and 
limits). For comparison, we overplot the observed spectral energy distribution
of Arp 220 scaled in flux and shifted to z=3 (small squares) and a schematic SED 
representing local ultraluminous galaxies, again shifted to z=3.}
\label{fig:sedplot}
\end{figure*}

\begin{figure*}
\center{\resizebox{17.9cm}{!}{\includegraphics{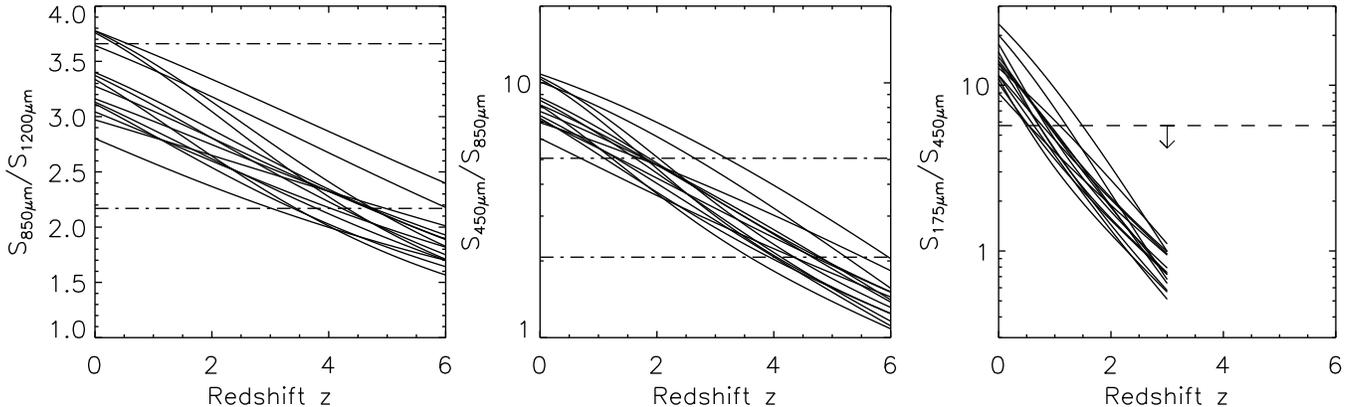}}}
\caption{Far-infrared photometric constraints on the redshift of Lockman850.1.
The continuous lines in each panel show the expected colors as a function
of redshift predicted from the SEDs of 14 local ULIRGs from Klaas et al. 
(\cite{klaas01}) that have
detailed far-infrared ISO photometry as well as submm detections.
Dashed lines indicate the allowed ranges or limits for Lockman850.1.}
\label{fig:klaasfits}
\end{figure*}

Table~\ref{tab:photom} summarizes the available photometry of Lockman850.1. 
The spectral energy distribution (SED) is shown in Fig.~\ref{fig:sedplot}, 
along with comparison SEDs for which we have chosen the observed SED  
of Arp220, which has an extremely steep rest frame mid- to far-IR spectrum,
and a schematic SED that is more typical for the properties 
of local ULIRGs. A first conclusion is that the SED of Lockman850.1 is well 
described by an ultraluminous (L$_{IR}\approx 8\,10^{12}$L$_\odot$ for 
H$_0$=75\,km\,s$^{-1}$\,Mpc$^{-1}$, q$_0$=0.5) infrared galaxy at at 
redshift of around 3.

The redshift of the source can be estimated on the basis
of its mm-to-radio spectral index. Using the `mean' relation for the 
350GHz/1.4GHz spectral index derived by Carilli \& Yun (\cite{carilli00}, 
see also http://www.aoc.noao.edu/\~\/ccarilli/alphaz.shtml),
the index of 0.90 observed for Lockman850.1 corresponds to a most probable
$z=2.7$.
Conservatively adopting their `z$_+$' relation and the 1$\sigma$ errors
on the radio and submm flux the lower limit is $z=1.4$. 
Taking into account the flattening of the mm-to-radio spectral index curve
for high redshifts and the variance of galaxy SEDs, the mm-to-radio index
of Lockman850.1 is consistent with all redshifts higher than these values
but probably not above $\approx$6 because of inverse Compton losses off the
cosmic microwave background (Carilli \& Yun \cite{carilli99}).  
A very high redshift is also excluded by the submm SED. We have used 
far-infrared and submm photometry of local ULIRGs by Klaas et al. 
(\cite{klaas01}) to assess constraints on the redshift of Lockman850.1
and the uncertainty introduced by SED variance. Figure~\ref{fig:klaasfits}
shows the expected colors for those 14 objects in their sample which have 
both detailed far-infrared photometry out to 200$\mu$m and at least one 
submm detection. Their
parametrization of the far-infrared/submm SEDs as single temperature 
modified blackbodies with finite optical depth is degenerate
with alternative multiple temperature parametrizations. This does not affect 
Fig.~\ref{fig:klaasfits}, however, which just requires a good fit to the data.
The predicted flux ratios shown in Fig.~\ref{fig:klaasfits} change with 
redshift as rest wavelengths move from the Rayleigh-Jeans part of a modified 
blackbody 
curve through and beyond its peak. The large 850$\mu$m to 1.2mm flux ratio 
is consistent with redshifts between 0 and $\approx$5 taking into acount both
the 1$\sigma$ flux errors and the 1$\sigma$ scatter in source redshifts for a 
given flux ratio. Recent observations of a fairly constant 
450$\mu$m to 850$\mu$m flux ratio for local galaxies (Dunne \& Eales
\cite{dunne01}) suggest that the parametrization chosen by Klaas et al.
(\cite{klaas01}) may overestimate the scatter in expected 850$\mu$m to 
1.2mm flux ratios for moderate redshift sources, but the flux errors 
and crosscalibration uncertainties remain a limitation for using this ratio.
The 450$\mu$m to 850$\mu$m ratio for Lockman850.1 suggests 
$z\approx 1.2\ldots5$, while the 175 to 450$\mu$m limit is not strongly
constraining ($z>0.6$).
Further limits on the redshift are placed by the nondetection of Lockman850.1 
by ISOCAM at 7 and 15$\mu$m.
For an SED as shown in Fig.~\ref{fig:sedplot} the
15$\mu$m to 850$\mu$m index implies $z>2.1$. The SED of Arp 220 with its very 
low mid- to far-infrared flux ratio, however, would imply $z>1.5$. Overall, 
the variance of observed infrared/submm/radio SEDs of galaxies allows
redshifts in the range 2 to 5 for Lockman850.1. This is in good agreement
with estimates we have obtained based on modelled SEDs (Efstathiou et al. 
\cite{efs00}) that suggest $z=2\ldots4$ with a best guess of $z=3$. 

The strong limit on the I band flux, classifying Lockman850.1 as an ERO,
shows that the rest-frame optical/UV SED of Lockman850.1 cannot be that of
a `blue' starburst. The SED plotted in Fig.~\ref{fig:sedplot} assumes a ratio
of infrared and blue luminosity representative for a ULIRG, but with a blue
optical/UV continuum (scaling SED SB6 from Kinney et al. 
\cite{kinney96}). This is clearly inconsistent with the I band limit.
An intrinsically blue optical/UV SED would be possible for a very high redshift 
I band dropout, but this
is inconsistent with the submm SED. Trentham et al. (\cite{trentham99})
have demonstrated that the slope of the optical/UV SEDs of local ULIRGs 
show large variations. A very steep, obscured SED like that
of VII\,Zw\,031 shifted to $z\sim3$ matches the observations of Lockman850.1. 

Luminous infrared galaxies may be powered either by star formation or an AGN.
The mid-infrared faintness of Lockman850.1 is not suggestive of a strong
AGN contribution. The XMM soft X-ray (rest frame hard X-ray) upper limit 
suggests 
L$_{2-10kev}$/L$_{bol}\ls 10^{-3}$ and excludes an unobscured type 1 AGN. 
In the submillimetre to X-ray spectral index vs. redshift diagram of 
Fabian et al. (\cite{fabian00}), the index limit of $>$1.4 places Lockman850.1
even beyond Compton-thick obscured AGN with properties similar to NGC 6240, for 
all plausible redshifts. 
If Lockman850.1 hosts an AGN, it must be weak and/or heavily obscured.
The absence of an obvious central AGN in the K image is consistent
with these SED arguments.

The K band morphology of Lockman850.1 is clumpy and extended over about
4\arcsec\/. It is clearly not that of
a relaxed passive elliptical at intermediate redshift, as often found
for (infrared-quiet) extremely red objects from near-IR surveys (Moriondo et al.
\cite{moriondo00}). On smaller scales, similar complexity has been 
observed in the R band image of another SCUBA source, SMM 14011+0252
(Smail et al. \cite{smail00a}).
For a range of redshifts (z=2$\ldots$5) and popular cosmologies
(H$_0$=65\,km\,s$^{-1}$\,Mpc$^{-1}$, ($\Omega_0,\Omega_\Lambda $) = (1,0) 
or (0.3,0.7)) 
1\arcsec$\approx\/4\ldots$8kpc, i.e. Lockman850.1 is a large system and
even the individual clumps approach galactic scales. This morphology is 
more complex than the interacting galaxies associated with some other
clearly identified (sub)mm sources (e.g. Smail et al. \cite{smail98}, 
Ivison et al. \cite{ivison00}, Bertoldi et al. \cite{bertoldi00}).
Interacting non-merged systems are usually characterized by two nuclei and
relatively faint tidal tails. More galaxies may be physically associated or 
projected, but the probability of obtaining a complex multi-peak morphology 
is small.
The distinction between galaxy interactions and patchy star formation
in a single large galaxy is a fuzzy one, of course, in the context of
hierarchical buildup of large galaxies from smaller units.
Lockman850.1 does not resemble an interacting system in the specific
meaning of an interaction of two well formed disk galaxies.

A nonnegligible fraction of bright submm sources is expected to be lensed 
(Blain et al. \cite{blain99}). There is no obvious candidate lens galaxy close to 
Lockman850.1, however, and there
are arguments against a model in which the brightest K peak
is the lens and the surrounding fainter clumps lensed images of the
background submm source. The submm emission then would be 
associated with the fainter clumps around the lens only, and thus
extended on a 4 arcsec scale without a central maximum. This is 
inconsistent with the
compactness of the interferometric image, and the good agreement of 
interferometric and large beam (sub)mm fluxes. In addition, the nondetection
in the I band implies that the I$-$K colors of all components must be extremely
red, i.e. there is no evidence for a color difference between putative 
`lens' and `lensed object'.
 
Most plausibly the clumps of Lockman850.1 are part of a large
and structured star forming galaxy, the clumps being 
individual very big star forming complexes and/or extinction minima in
a patchy obscuration pattern seen against the rest frame optical/UV
emission of the source. It has been argued on the basis of the
huge inferred star formation rates ($\approx$1000\,M$_\odot$/yr depending
on assumptions about dust temperature and initial mass function) that
submm galaxies highlight the formation of massive spheroids -- these are
exactly the star formation rates required to build a massive spheroid
in a short time ($\sim$1\,Gyr) if the formation is in one localised starburst. 
This scenario is strengthened by the consistency 
between the current mass in spheroids
and the one inferred from IR/Submm counts (Tan et al. \cite{tan99}).
Lockman850.1 may further support this by its morphology: 
The complex K band morphology as well as the K magnitude are reminiscent 
of $z>3$ radio galaxies (van Breugel et al. \cite{vanbreugel98}, see also
Pentericci et al. \cite{pente01}), and radio
galaxies in the local universe are known to reside in massive ellipticals.
Strong dust emission observed in many high-z radio galaxies (Archibald et al.
\cite{archibald01}) is further evidence for similarities between submm
sources and high-z radio galaxies. Lockman850.1 is also very luminous in the
rest frame optical: adopting the estimated $z\approx 3$, 
it must be at least as bright as M(B$_{rest}$)$\approx -25$. 
Even taking into account evolutionary corrections, this suggests a very 
massive system. If the morphology of Lockman850.1 reflects the
buildup of an elliptical from $\approx$10\,kpc size clumps, observations
of submm sources may constrain the cosmic time of these events in a
hierarchical merger picture. 
Submm surveys may highlight the formation of some of today's massive 
ellipticals.

\begin{acknowledgements}
We are indebted to Klaus Meisenheimer for providing access to a 
CADIS K-band image of the Lockman hole region, and to Ulrich Klaas and
Martin Haas for access to their ULIRG photometry. We thank Dario Fadda
and Paolo Ciliegi for information on unpublished mid-infrared and radio 
data. Special thanks go to the IRAM
staff for excellent support during a difficult period.
\end{acknowledgements}

\end{document}